# GeoCode-GPT: A Large Language Model for Geospatial Code Generation Tasks

Shuyang Hou[a], Zhangxiao Shen[a], Anqi Zhao[a], Jianyuan Liang[a], Zhipeng Gui[b], Xuefeng Guan[a], Rui Li[a], Huayi Wu[a]*

*a. State Key Laboratory of Information Engineering in Surveying, Mapping and Remote Sensing, Wuhan University, Wuhan, China

b. School of Remote Sensing and Information Engineering, Wuhan University

Wuhan, China

*Corresponding author: Huayi Wu, email: wuhuayi@whu.edu.cn

**Abstract**

The increasing demand for spatiotemporal data and modeling tasks in geosciences has made geospatial code generation technology a critical factor in enhancing productivity. Although large language models (LLMs) have demonstrated potential in code generation tasks, they often encounter issues such as refusal to code or hallucination in geospatial code generation due to a lack of domain-specific knowledge and code corpora. To address these challenges, this paper presents and open-sources the GeoCode-PT and GeoCode-SFT corpora, along with the GeoCode-Eval evaluation dataset. Additionally, by leveraging QLoRA and LoRA for pretraining and fine-tuning, we introduce GeoCode-GPT-7B, the first LLM focused on geospatial code generation, fine-tuned from Code Llama-7B. Furthermore, we establish a comprehensive geospatial code evaluation framework, incorporating option matching, expert validation, and prompt engineering scoring for LLMs, and systematically evaluate GeoCode-GPT-7B using the GeoCode-Eval dataset. Experimental results show that GeoCode-GPT outperforms other models in multiple-choice accuracy by 9.1% to 32.1%, in code summarization ability by 1.7% to 25.4%, and in code generation capability by 1.2% to 25.1%. This paper provides a solution and empirical validation for enhancing LLMs' performance in geospatial code generation, extends the boundaries of domain-specific model applications, and offers valuable insights into unlocking their potential in geospatial code generation.

## 1. Introduction

Code generation technology translates natural language into source code (NL2Code), which can pertain to general-purpose programming languages such as Python, C, or JavaScript, or domain-specific languages for specialized platforms like MATLAB, Bioconductor, or Google Earth Engine(Jiang et al., 2024). Geospatial code is specifically designed for geospatial analysis platforms (e.g., Google Earth Engine, ArcGIS) or modeling tasks, typically implemented through wrappers around general-purpose languages or by invoking specialized geocomputational tools and libraries, such as GDAL and ArcPy in Python, or the Raster and Terra packages in R(Coetzee et al., 2020). With the surge in spatiotemporal data, the growing demand for geospatial modeling tasks has made geospatial code generation technology a critical factor in enhancing productivity(Zhang et al., 2024b).

Code generation is fundamentally a subset of natural language processing (NLP). Early research in this area primarily relied on heuristic rules(Nymeyer and Katoen, 1997) and expert systems(Lirov, 1991), such as probabilistic grammar-based frameworks(Kifetew et al., 2017) and small-scale pretrained models(Li et al., 2019). These methods depended on annotated data, had limited applicability, and were difficult to scale. With the advent of large language models (LLMs) based on self-attention mechanisms and the Transformer architecture, models trained on vast corpora have demonstrated "emergent" abilities, such as task execution, in-context learning, and reasoning(Zhao et al., 2023). General-purpose LLMs specifically designed for code generation, such as Code LLaMA(Roziere et al., 2023), DeepSeek Coder(Guo et al., 2024), and WizardCoder(Luo et al., 2023), are reshaping the research paradigm in this field. These models' instruction-based code generation capabilities allow even programming novices to produce functional code, driving the democratization of programming and significantly boosting productivity, enabling users to focus on higher-level logic and planning.

General-purpose code generation LLMs are predominantly trained on code related to standard programming tasks. In contrast, geospatial code typically handles complex spatiotemporal data, characterized by specific data formats (e.g., geographic coordinates, multi-dimensional rasters, multi-band spectra) and massive datasets (e.g., global remote sensing collections)(Hou et al., 2024). Since geospatial code is often executed on specialized platforms with unique function syntax and data flow control logic, it significantly differs from general-purpose code(Granell et al., 2010). Furthermore, these platforms' built-in datasets frequently use proprietary internal indexing and naming conventions, which general LLMs are usually unfamiliar with. As a result, when generating geospatial code, models may fail to encode correctly, or produce code with incorrect operator choices, non-compliant syntax, unreasonable parameters, incomplete logic, or mismatched inputs and outputs. These issues, illustrated in Fig. 1, are referred to as "refusal to code" or "coding hallucinations(Hou et al., 2024)."

In response to the challenges posed by the lack of domain-specific corpora and dedicated models for geospatial code generation in general LLMs, this study introduces the GeoCode-PT pretraining corpus, the GeoCode-SFT supervised fine-tuning corpus, and the GeoCode-Eval evaluation dataset. GeoCode-PT aggregates multi-source data closely related to geospatial code, including specialized code from platforms like Google Earth Engine and PIE Engine, as well as geocomputational tool code from platforms such as ArcGIS, Matlab, RStudio, and Python. It comprises a total of 275,374 code snippets, 10,190 operator knowledge entries, 853 dataset knowledge entries, and 14 common knowledge documents on platforms and toolsets. GeoCode-SFT, built using a structured traversal algorithm and the Self-Instruct framework, generated 502,047 instruction data entries, covering operators, datasets, platform language understanding, and code summarization. GeoCode-Eval consists of 3,000 multiple-choice questions, 500 code generation tasks, and 500 code summarization tasks.

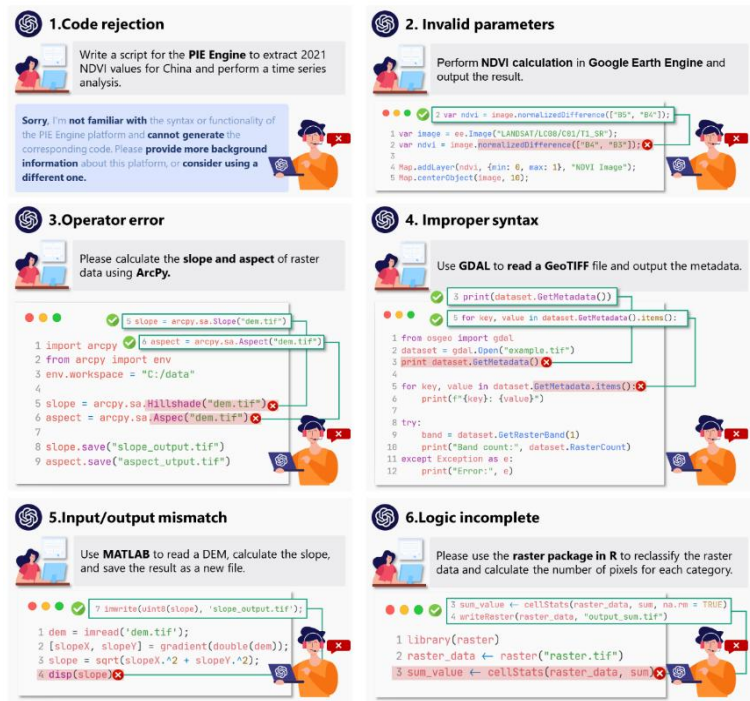

Fig.1. Illustration of Refusal to Code or Coding Hallucinations

Based on GeoCode-PT and GeoCode-SFT, this study employs an autoregressive unsupervised approach to conduct lightweight pretraining on Code Llama-7B using QLoRA, followed by supervised fine-tuning with instruction-tuning data via LoRA. This process culminated in the release of GeoCode-GPT-7B, the first LLM dedicated to geospatial code generation. Subsequently, GeoCode-GPT underwent comprehensive evaluation using GeoCode-Eval, across eight key dimensions: Operator Knowledge, Dataset Knowledge, Platform or Toolkit Knowledge, Platform or Toolkit Recognition, Programming Language Recognition, Entity Recognition, Code Summarization, and Code Generation. The evaluation was conducted through option matching, expert validation, and prompt engineering scoring. With the enrichment of extensive geospatial knowledge, GeoCode-GPT outperforms general-purpose models in several capabilities, and despite having significantly fewer parameters than commercial models, it approaches their performance in certain metrics. This achievement advances the application and development of LLMs in geospatial code generation. The overall research framework is illustrated in Fig. 2.

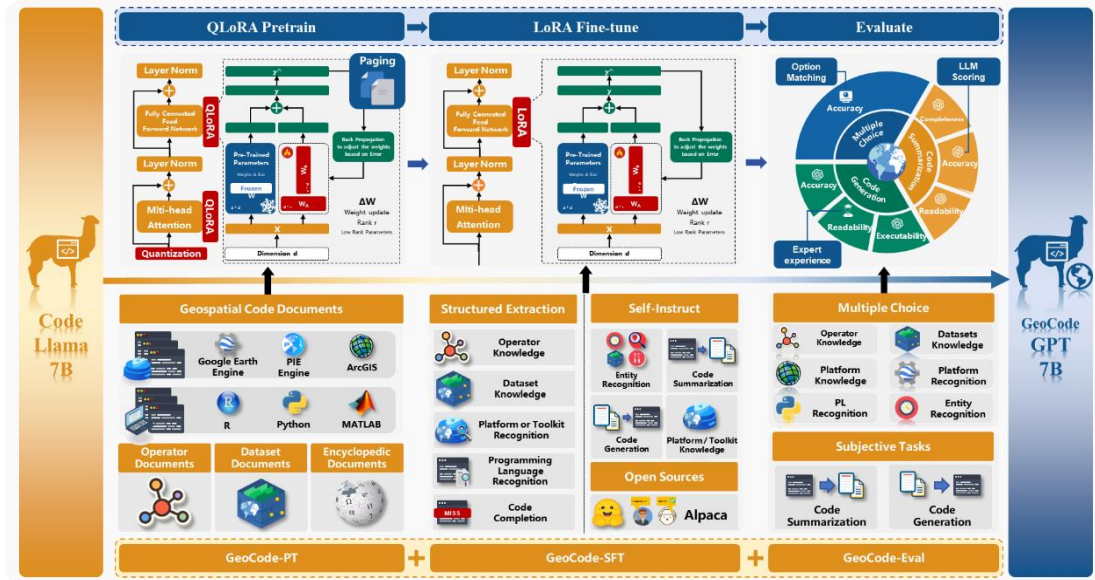

Fig.2. Overall Framework of GeoCode-GPT

Our contributions are summarized as follows:
- We introduce GeoCode-GPT-7B, the first LLM specifically designed for geospatial code generation tasks.
- We developed and open-sourced the GeoCode-PT and GeoCode-SFT corpora, as well as the GeoCode-Eval evaluation set, providing a systematic corpus foundation and evaluation tools for pretraining and fine-tuning LLMs in geospatial code generation tasks. https://github.com/whuhsy/GeoCode-GPT.
- We propose a novel pretraining and fine-tuning strategy that combines QLoRA and LoRA, offering a new approach to balancing computational resources and training efficiency.
- We established a comprehensive evaluation framework for geospatial code, incorporating option matching, expert validation, and prompt engineering scoring for LLMs, providing insights for the holistic evaluation of domain-specific models.

The remainder of this paper is organized as follows: Section 2 reviews relevant research on the specialization of LLMs in geosciences, code generation, fine-tuning data generation, parameter-efficient fine-tuning (PEFT), and code generation evaluation. Section 3 describes the construction methods for the GeoCode-PT, GeoCode-SFT, and GeoCode-Eval corpora. Section 4 details the selection of the base model for training, along with the parameter settings for pretraining and fine-tuning. Section 5 discusses the evaluation dimensions, methods, and result analysis based on GeoCode-Eval. Finally, Section 6 summarizes the contributions of GeoCode-GPT, analyzes its limitations, and outlines future research directions.

## 2. Related Work

### 2.1 Challenges in Domain Specialization for LLMs

LLMs, based on the Transformer architecture and self-attention mechanisms, exhibit remarkable NLP capabilities due to their large-scale parameters and diverse training corpora(Minaee et al., 2024). However, they face two major challenges in domain adaptation: first, retraining requires significant computational resources, making it difficult to rapidly update domain-specific knowledge, which can lead to knowledge gaps(Kasneci et al., 2023). Second, the limited availability of specialized data in general-purpose corpora may result in "knowledge hallucinations(Tonmoy et al., 2024)," where the model generates content that appears correct but is factually inaccurate in highly specialized tasks. Thus, advancing the domain specialization of LLMs has become a key research focus, with the goal of enhancing model performance by integrating domain-specific knowledge. While there have been successful applications in fields such as law(Cui et al., 2023), finance(Huang et al., 2023; Wu et al., 2023), biomedicine(Thirunavukarasu et al., 2023), and earth sciences(Zhang et al., 2024a), research on domain-specific LLMs for geospatial code generation remains unexplored.

### 2.2 Instruction Data Generation and Fine-Tuning Strategies for Code Generation in LLMs

The application of LLMs in code generation relies on their capabilities in instruction execution, contextual learning, and reasoning to transform natural language descriptions into source code. High-quality instruction datasets are crucial for enhancing a model's ability to follow instructions. Research has shown that LIMA-65B, fine-tuned with just 1,000 carefully designed prompts and responses, achieved comparable performance to GPT-4 in 43% of evaluation cases(Zhou et al., 2023). However, obtaining high-quality data is challenging due to issues of scarcity, privacy, and cost. The Self-Instruct framework addresses these limitations by enabling models to generate synthetic data autonomously(Wang et al., 2022), eliminating the need for manual annotation. This approach has been successfully applied in models like Alpaca(Zhang et al., 2023) and Phi(Abdin et al., 2024), demonstrating the effectiveness of synthetic data in training. Models are first pretrained on large-scale unannotated code to grasp structural and semantic patterns, followed by supervised fine-tuning through instruction tuning. While full fine-tuning (FFT) yields significant results, it is highly resource-intensive. PEFT methods, such as LoRA(Hu et al., 2021), reduce the computational burden by minimizing parameter updates, and QLoRA further enhances fine-tuning efficiency in resource-constrained settings through int4 quantization(Qin et al., 2024). Although LLMs have made notable progress in the field of code generation, particularly with open-source models like Meta AI's Code LLaMA, DeepSeek Coder, and WizardCoder, there remains a lack of specialized pretrained and instruction datasets for geospatial code generation. Moreover, fine-tuning strategies tailored for this domain have yet to be applied, and no dedicated LLM product for geospatial code generation has emerged.

## 2.3 Evaluation of Code Generation Quality in LLMs

In the task of code generation for LLMs, traditional token-matching metrics such as Exact Match, BLEU, and ROUGE struggle to accurately assess code executability, syntactic correctness, and functionality(Chen et al., 2024). To address this, execution-based metrics are increasingly adopted, such as pass@k, which evaluates the probability that at least one of the k generated code samples passes all unit tests, offering a more reliable reflection of the actual performance of the generated code(Chen et al., 2021). However, geospatial code often involves complex image processing and relies on platform-specific operations, making automated evaluation more challenging. Moreover, code evaluation must also consider factors such as readability, conciseness, and the accuracy of data usage, further limiting the comprehensiveness of automated assessments(Scalabrino et al., 2019). While human evaluation remains important for validating code quality, it is hindered by subjectivity and inconsistent standards. In recent years, methods leveraging LLMs for automatic evaluation have gained traction. By designing specific prompts, LLMs can act as "judges" (LLM-as-a-Judge), as demonstrated in projects like AlpacaEval and MT-bench(Hui et al., 2024). In these cases, LLMs not only provide scores but also explain the rationale behind their evaluations, improving interpretability. However, the application of this approach to code generation is still in its early stages. The ICE-Score, which assesses code functionality and human preference through LLM evaluations, has shown promising correlation with human judgments and does not require additional benchmarks or references(Zhuo, 2023). This provides a novel evaluation approach, but it has yet to be applied to the evaluation of geospatial code generation.

## 3. Dataset Construction

This section provides a detailed description of the construction methods and processes for the pretraining corpus (GeoCode-PT), the supervised fine-tuning corpus (GeoCode-SFT), and the evaluation corpus (GeoCode-Eval) used in training the GeoCode-GPT model.

## 3.1 GeoCode-PT

In the pretraining phase, an unsupervised learning approach is employed, where the model autonomously learns code structures and semantic relationships by processing large-scale unannotated data, without relying on explicit instructions or labeled data. The pretraining corpus, GeoCode-PT, is designed to provide the model with broad foundational knowledge in the domain. A detailed breakdown of the GeoCode-PT dataset is shown in Table 1, comprising four main components: code documentation, operator knowledge documents, dataset knowledge documents, and encyclopedic documents.

Table 1 GeoCode-PT Data Inventory

| Training Dataset | Platform/Library | Language | Format | Quantity | Size |
|---|---|---|---|---|---|
| Code Documents | Google Earth Engine | JavaScript | JSON | 128396 | 5.57GB |
| | | Python | JSON | 107826 | |

| Training Dataset | Platform/Library | Language | Format | Quantity | Size |
|---|---|---|---|---|---|
| | ArcPy | Python | JSON | 30122 | |
| | PIE Engine | JavaScript | JSON | 1198 | |
| | Mapping Toolbox | Matlab | JSON | 3218 | |
| | GDAL | Python | JSON | 1310 | |
| | rasterio | Python | JSON | 86 | |
| | cartopy | Python | JSON | 257 | |
| | GeoPandas | Python | JSON | 197 | |
| | sf | R | JSON | 488 | |
| | terra | R | JSON | 1952 | |
| | gstat | R | JSON | 324 | |
| | Overall | | | 275374 | |
| | Google Earth Engine | JavaScript | csv | 1374 | |
| | | Python | csv | 1348 | |
| | ArcPy | Python | JSON | 2025 | |
| | PIE Engine | JavaScript | JSON | 962 | |
| | Mapping Toolbox | Matlab | JSON | 317 | |
| **Operator Documents** | GDAL | Python | JSON | 2073 | 13.48MB |
| | rasterio | Python | JSON | 1538 | |
| | cartopy | Python | JSON | 16 | |
| | GeoPandas | Python | JSON | 170 | |
| | sf | R | JSON | 85 | |
| | terra | R | JSON | 240 | |
| | gstat | R | JSON | 42 | |
| | Overall | | | 10190 | |
| **Dataset Documents** | GEE | / | csv | 645 | |
| | PIE Engine | / | csv | 208 | 309KB |
| | Overall | | | 853 | |
| **Encyclopedic Documents** | Wikipedia | Natural Lang. | JSON | 4 | |
| | Built-in Platform Docs | Natural Lang. | JSON | 10 | 102KB |
| | Overall | | | 14 | |

The code documentation includes specialized code from cloud platforms such as Google Earth Engine and PIE Engine, as well as local code written with geospatial computation tools specific to platforms like ArcGIS, Matlab, RStudio, and Python. A total of 275,374 code snippets were collected from platforms like GitHub, Stack Overflow, and Hugging Face, screened for syntax accuracy, and preserved with comment information. The operator knowledge documents compile commonly used operators and their syntactic details from various geoscience platforms, sourced from official documentation. These entries, totaling 10,190, include the full name, abbreviation, description, output type, and parameters. The dataset knowledge documents provide detailed information on frequently used built-in datasets from platforms like Google Earth Engine and PIE Engine, also sourced from official pages. This section contains 853 entries, including unique identifiers, names, platforms, directories, tags, DOIs, and related links. Additionally, 14 encyclopedic documents from Wikipedia offer overviews of various geospatial analysis platforms and their tools. The dimensional information of each document category is summarized in Table 2.

Table 2 GeoCode-PT Data Attribute Table

| Type | Attribute |
|---|---|
| Code documents | Code_id |
|  | Language |
|  | Platform |
|  | Library |
|  | Title |
|  | Description |
|  | Content |
| Dataset documents | Dataset_id |
|  | Name |
|  | Provide |
|  | Snippet |
|  | Tags |
|  | Description |
|  | DOI |
|  | Website |
| Operator documents | Operator_id |
|  | Full_name |
|  | Short_name |
|  | Library_name |
|  | Language |
|  | Platform |
|  | Description |
|  | Usage |
|  | Parameters |
|  | Output_type |
| Encyclopedic Documents | Name |
|  | Text |

## 3.2 GeoCode-SFT

The supervised fine-tuning corpus consists of high-quality, domain-specific instructions following an "Instruct-Input-Output" structure, where the "Instruct" prompts the task, the "Input" provides the input data, and the "Output" generates the correct answer(Wang et al., 2024). These instructions are designed to enhance the model's ability to generate geospatial code, ensuring that GeoCode-GPT can accurately interpret human instructions and produce natural language outputs. GeoCode-SFT includes both geospatial code-related instructions and natural language-based instructions. The data breakdown is presented in Table 3. The construction of GeoCode-SFT involves the acquisition of open-source data, structured extraction, and Self-Instruct generation based on LLMs.

Table 3 GeoCode-SFT Data Inventory

| Production Method | Dataset Type | Quantity |
|---|---|---|
| Structured | Operator Knowledge | 26,130 |
|  | Dataset Knowledge | 2,511 |
|  | Platform or Toolkit Recognition | 18,827 |
|  | Programming Language Recognition | 18,827 |
| Large Model | Code Completion | 14,817 |
|  | Entity Recognition | 18,827 |
|  | Code Summarization | 18,827 |
|  | Code Generation | 18,827 |
|  | Platform or Toolkit Knowledge | 454 |
| Open Source | Alpaca | 7*52,000 |

### 3.2.1 Structured Extraction

Structured information extraction is primarily achieved through a rule-based traversal algorithm to generate instruction sets, as illustrated in Fig. 3. The first method, rule slicing, is applicable to structured data formats such as JSON and CSV. By setting rules to partition the data, instruction sets containing "instruct," "input," and "output" are generated. For instance, in CSV data, the horizontal axis represents subjects, and the vertical axis represents attributes. By matching these, the attribute values for the subject are extracted, with "instruct" as the task description, "input" as the subject name, and "output" as the attribute value. This algorithm can quickly generate large volumes of instruction data, as shown in Fig. 3-(a). In JSON data, the key represents the attribute, and the value represents the data, which, when matched, forms the "input" and "output" fields, as shown in Fig. 3-(b).The second method, rule masking, is used for completing text or code snippets. For example, a code snippet is divided into three parts—prefix, middle, and suffix—based on sentence count. Through masking, "instruct" describes the completion task, "input" is the incomplete code snippet, and "output" is the completion portion, forming the instruction data, as shown in Fig. 3-(c).

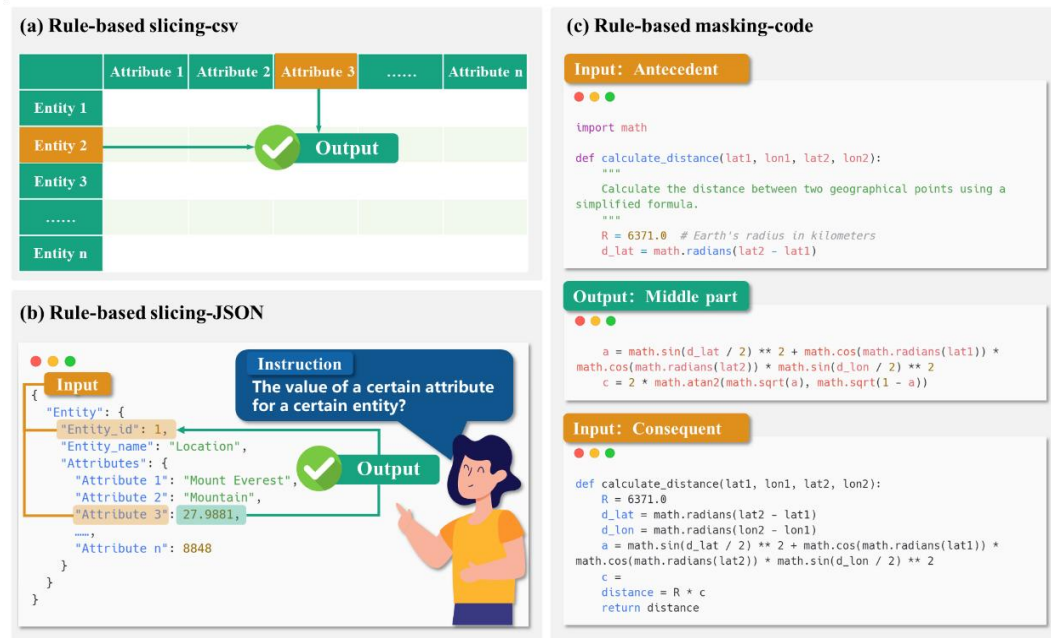

Fig.3. Diagram of the Rule-based Structured Instruction Construction Method

### 3.2.2 Self-Instruct generation

The instruction tuning data from structured extraction primarily organizes the textual content of the pretraining data in sequence, functioning as a form of "pre-disclosure" for downstream tasks. This method emphasizes surface-level information organization but lacks deeper semantic connections. In contrast, the Self-Instruct framework generates synthetic instruction sets through the model's reasoning abilities, uncovering the underlying semantic relationships and logical patterns. This approach not only systematically organizes knowledge but also reveals deeper logical relationships, aiming to significantly enhance the model's understanding and application capabilities in geospatial code generation tasks. The construction process is illustrated in Fig. 4.

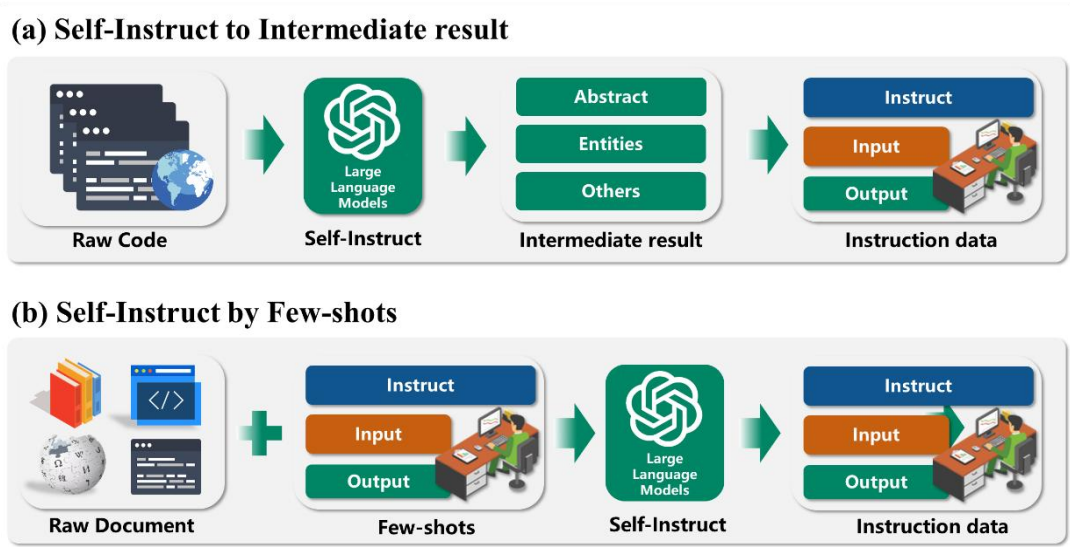

Fig.4 Instruction Data Construction Method Based on the Self-Instruct Framework

The core of the Self-Instruct framework lies in guiding the large language model to self-summarize through prompt engineering, thereby generating task-relevant instruction sets. This can be achieved in two ways: first, after producing intermediate results, instruction sets are constructed based on rules and machine learning algorithms, as illustrated in Fig. 4-(a). For example, high-quality code is selected from GeoCode-PT to generate code summaries, and instructions are created through bidirectional pairing of the code and its summary.The second approach involves one-shot learning, where a few examples are directly provided, allowing the model to autonomously summarize knowledge and generate instructions, as shown in Fig. 4-(b). For instance, extracting relationships between operators and datasets from knowledge tables, or generating platform and toolkit recognition instructions from encyclopedic documents.

### 3.2.3 Natural Language Style Instruction Tuning Data

To prevent pure code-related instruction data from potentially diminishing the general language understanding and coding capabilities of the model during fine-tuning(Zhang et al., 2024a), we collected the multilingual code understanding and generation instruction dataset, Alpaca-GPT-4, from the Hugging Face platform. We filtered the dataset to include multilingual data such as Chinese, English, Japanese, Arabic, and French. These diverse natural language instructions help enhance GeoCode-GPT's multilingual adaptability and improve the readability of its outputs.

### 3.3 GeoCode-Eval

The evaluation corpus is designed to test the model's performance in geospatial code generation tasks. Based on Benjamin Bloom's taxonomy of cognitive objectives(Arievitch, 2020), we developed the GeoCode-Eval dataset, which covers three key dimensions: "Cognition and Memory," "Comprehension and Interpretation," and "Innovation and Creation," as illustrated in Fig. 5.

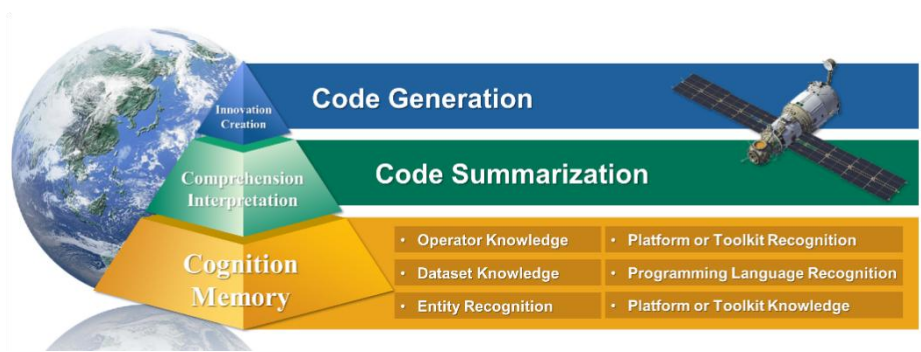

Fig.5 GeoCode-Eval Evaluation Dimension Construction Diagram

In the "Cognition and Memory" phase, GeoCode-Eval assesses the model's basic cognitive abilities through multiple-choice questions, including operator knowledge, dataset knowledge, platform recognition, and entity recognition. In the Comprehension and Interpretation phase, the model's ability to summarize is examined, evaluating its capacity to synthesize and interpret geospatial code. Finally, in the Innovation and Creation phase, the model's code generation capability is assessed through code generation tasks. The detailed data breakdown is presented in Table 4.

Table 4 GeoCode-Eval Data Inventory

| Type | Evaluation Dataset | Quantity | Creation Method | Requirements |
|---|---|---|---|---|
| **Multiple Choice** | Operator Knowledge | 1000 | Opensource Or Self-Instruct | Uniqueness Accuracy Deceptiveness |
| | Dataset Knowledge | 250 | | |
| | Platform or Toolkit Knowledge | 250 | | |
| | Platform or Toolkit Recognition | 500 | | |
| | Programming Language Recognition | 500 | | |
| | Entity Recognition | 500 | | |
| **Subjective** | Code Summarization | 500 | Prompt Engineering | Accuracy Clarity Readability |
| | Code Generation | 500 | | |

To ensure fairness in evaluation and prevent knowledge leakage, all multiple-choice questions in GeoCode-Eval were sourced from platforms such as Baidu Wenku, Niuke, and Daoke Wenku, covering university exams and basic programming tests, or generated using the Self-Instruct framework based on GeoCode-PT data. All questions were manually validated to ensure the uniqueness of the correct answers and the plausibility of the distractors, resulting in a total of 3,000 multiple-choice questions. The summarization and code generation tasks were constructed similarly to GeoCode-SFT, using different valid code snippets, with prompt-based guidance for generating geospatial code summaries. The summaries were required to accurately capture functionality, clearly identify entities, and be concise in language. A total of 500 summarization tasks and 500 code generation tasks were created, with all questions and answers manually checked for quality and readability.

## 4. Model Training

A high-quality base model is crucial for effective fine-tuning. Compared to general-purpose LLMs, general-purpose code generation models are already optimized for code generation tasks but lack domain-specific knowledge of geospatial code. Therefore, this study focuses on fine-tuning a general code generation model. QLoRA reduces GPU memory usage by approximately 75%, enabling larger batch sizes and longer sequences, making it suitable for large-scale code processing in resource-constrained environments. LoRA, on the other hand, accelerates convergence during fine-tuning by about 66%, making it ideal for high-precision instruction tuning. As a result, we adopt QLoRA for pretraining and LoRA for fine-tuning to achieve an optimal balance between computational resources and training efficiency. This section introduces the base model, Code Llama-7B, and the two training stages of GeoCode-GPT.

### 4.1 Base Model

In this study, we selected Code Llama as the base model. LLaMA-2 is a commonly used foundation for open-source code generation models, and Code Llama builds on this foundation with specialized training. It has demonstrated excellent performance in benchmark tests such as HumanEval and MBPP, particularly in tasks like code generation, code completion, and handling long contexts, even surpassing the larger LLaMA 2-70B model in some areas. This makes it highly suitable for optimizing geospatial code generation tasks. Additionally, Code Llama offers three parameter configurations: 7B, 13B, and 34B. Compared to the larger models, Code Llama-7B strikes an ideal balance between performance and lower resource consumption, while also supporting efficient LoRA fine-tuning. Considering performance, flexibility, and training costs, Code Llama-7B is the optimal choice for this study.

### 4.2 Pretraining

During the pretraining phase, we optimized the Code Llama-7B model using the GeoCode-PT corpus with QLoRA, combining quantization and low-rank adaptation to reduce memory usage and support the longest possible input sequences, thereby improving the model's learning efficiency. The specific hyperparameter configuration is as follows: a global batch size of 64, int4 quantization precision, and one training epoch. The initial learning rate was set at 0.0002, with a cosine decay scheduler and linear warmup over the first 5% of steps. A weight decay factor of 0.1 was applied to prevent overfitting. Gradient accumulation steps were set to 4, and the maximum input sequence length was 4096 tokens, ensuring smooth operation on two NVIDIA A100 40GB GPUs.For the QLoRA configuration, we set the rank value to 64, used NF4 quantization, and a scaling factor of 128. The main modules optimized included the q_proj, v_proj, k_proj, o_proj, and MLP layers. A dropout probability of 0.05 was applied to mitigate the risk of overfitting.

### 4.3 Fine-Tuning

After QLoRA pretraining, the model acquired foundational knowledge in geospatial code generation. To further enhance the model's controllability and generalization to new tasks, we applied LoRA for lightweight fine-tuning, enabling more precise parameter adjustments. The fine-tuning settings were as follows: an initial learning rate of 0.0001 was chosen to ensure stability and gradual convergence during the fine-tuning process. The global batch size was set to 32, with gradient accumulation steps of 4, resulting in an effective batch size of 128. The input sequence length remained at 4096 tokens to handle longer contextual information.The LoRA configuration included a rank value of 64, targeting the key modules q_proj, v_proj, k_proj, o_proj, and the MLP. A dropout rate of 0.05 was maintained to prevent overfitting. The fine-tuning process was conducted on two NVIDIA A100 40GB GPUs, ensuring efficient utilization of computational resources and improving the model's performance in following complex instructions.

## 5. Evaluation

The evaluation was conducted using the GeoCode-Eval dataset. This section presents the evaluation methodology and results for GeoCode-GPT-7B, and compares its performance with leading LLMs across various domains. Commercial models included GPT-4, GPT-3.5, and ERNIE 4.0. For general open-source LLMs, we selected LLaMA 2-7B, LLaMA 3-8B, as well as general-purpose code generation models with similar parameter sizes: CodeGemma-7B, StarCoder 2-7B, and CodeGeeX 2-6B. The baseline model was Code Llama-7B, with the higher-parameter version, Code Llama-13B, included for reference.

### 5.1 Multiple-Choice Questions

Each multiple-choice question provides four options, with one correct answer, covering six dimensions: Operator Knowledge (OK), Datasets Knowledge (DK), Platform or Toolkits Knowledge (PTK), Platform or Toolkits Recognition (PTR), Programming Language Recognition (PLR), and Entity Recognition (ER). The model's output was then compared with the correct answer, and the corresponding accuracy was calculated. The evaluation results are presented in Table 5.

Table 5 : Multiple-choice Question Evaluation Data Table

| Model | Acc-OK | | Acc-DK | | Acc-PTK | | Acc-PTR | | Acc-PLR | | Acc-ER | | Acc-Average | |
|---|---|---|---|---|---|---|---|---|---|---|---|---|---|---|
| GPT-4 | 0.804 | +0.069 | 0.520 | +0.272 | 0.784 | -0.032 | 0.878 | +0.024 | 0.936 | +0.018 | 0.852 | -0.106 | 0.757 | +0.091 |
| GPT-3.5 | 0.752 | +0.121 | 0.452 | +0.340 | 0.732 | +0.020 | 0.852 | +0.050 | 0.914 | +0.040 | 0.728 | +0.108 | 0.738 | +0.110 |
| ERNIE 4.0 | 0.723 | +0.150 | 0.404 | +0.388 | 0.608 | +0.144 | 0.804 | +0.098 | 0.890 | +0.064 | 0.678 | +0.158 | 0.685 | +0.163 |
| LLaMA 2-7B | 0.476 | +0.397 | 0.328 | +0.464 | 0.640 | +0.112 | 0.676 | +0.226 | 0.822 | +0.132 | 0.524 | +0.312 | 0.578 | +0.270 |
| LLaMA 3-8B | 0.527 | +0.346 | 0.376 | +0.416 | 0.652 | +0.100 | 0.690 | +0.212 | 0.850 | +0.104 | 0.536 | +0.300 | 0.606 | +0.242 |
| CodeGemma-7B | 0.583 | +0.290 | 0.208 | +0.584 | 0.520 | +0.232 | 0.608 | +0.294 | 0.876 | +0.078 | 0.468 | +0.368 | 0.544 | +0.304 |
| StarCoder 2-7B | 0.597 | +0.276 | 0.272 | +0.520 | 0.572 | +0.180 | 0.610 | +0.292 | 0.880 | +0.074 | 0.316 | +0.520 | 0.541 | +0.307 |
| CodeGeeX 2-6B | 0.548 | +0.325 | 0.252 | +0.540 | 0.440 | +0.312 | 0.586 | +0.316 | 0.846 | +0.108 | 0.492 | +0.344 | 0.527 | +0.321 |
| Code Llama-13B | 0.605 | +0.268 | 0.312 | +0.480 | 0.600 | +0.152 | 0.636 | +0.266 | 0.902 | +0.052 | 0.451 | +0.385 | 0.584 | +0.264 |
| Code Llama-7B | 0.643 | +0.230 | 0.284 | +0.508 | 0.592 | +0.160 | 0.614 | +0.288 | 0.874 | +0.080 | 0.424 | +0.412 | 0.572 | +0.276 |
| **GeoCode-GPT-7B** | **0.873** | | **0.792** | | **0.752** | | **0.902** | | **0.954** | | **0.746** | | **0.848** | |

The evaluation results show that GeoCode-GPT-7B outperformed the baseline model, Code Llama-7B, across all assessed dimensions. The smallest improvement was observed in the Programming Language Recognition dimension (0.08), while the largest improvement was in the Datasets Knowledge dimension (0.508), with an overall average increase of 0.276. Compared to other models, GeoCode-GPT-7B scored 0.032 and 0.106 lower than GPT-4 in the Platform or Toolkits Knowledge and Entity Recognition dimensions, respectively, but demonstrated significant superiority in all other dimensions.

Since Operator Knowledge and Datasets Knowledge pertain to the "proprietary knowledge"

of geospatial platforms, they were not included in the training data of other models, leading to weaker performance in the Datasets Knowledge dimension. The lowest scores were 0.476 for LLaMA 2-7B and 0.208 for CodeGemma-7B. In contrast, after domain-specific fine-tuning, GeoCode-GPT-7B's scores increased by 0.066 to 0.397 (Operator Knowledge) and by 0.272 to 0.584 (Datasets Knowledge).In the Programming Language Recognition task, since geospatial code is wrapped around foundational programming languages, retaining some of the original syntax, most models could easily recognize the underlying languages. Commercial models performed the best, with general-purpose code generation models slightly outperforming general LLMs. GeoCode-GPT-7B showed limited improvement in this task (0.018-0.132).By comparison, the models performed worse in the Platform or Toolkits Recognition task, likely due to the lower usage of certain platforms and the limited availability of open-source code. While models could identify syntax differences, they struggled to accurately match platform or toolkit names without specific instruction training. However, GeoCode-GPT-7B showed significant improvement in this task due to instruction tuning, outperforming other models by 0.024 to 0.316 points.Platform or Toolkits Knowledge is generally considered common encyclopedic knowledge, resulting in minimal performance differences across models. Commercial models overall outperformed general LLMs, while general-purpose code generation models performed the worst, possibly because their training focuses more on code content, reducing their geospatial foundational knowledge. In this task, GeoCode-GPT-7B led by 0.02 to 0.312 points, except for GPT-4.In the Entity Recognition task, where code functions as a specialized form of natural language, stronger semantic understanding is required. As a result, general-purpose LLMs slightly outperformed general code generation models. GeoCode-GPT-7B demonstrated higher accuracy than other models (except GPT-4), with a margin of 0.108 to 0.52 points. These evaluation results validate the effectiveness of GeoCode-GPT-7B's training strategy.

**5.2 Subjective Questions**

Geospatial code often involves the analysis and processing of spatiotemporal imagery, which is difficult for standard compilers to directly handle or visualize. As a result, evaluation typically relies on experts, which is costly and subjective. Similar challenges arise when assessing the readability, entity accuracy, and completeness of code or summaries. Since the semantic expression of the same entity can vary, string-matching scores lack flexibility. To address this, we designed specific prompts for GPT-4 to act as an "evaluator" for automatic assessment. Given the token cost of LLMs, the prompt template was carefully crafted to emphasize conciseness in scoring, aiming to improve expression efficiency and avoid overly lengthy responses.

**5.2.1 Code Summarization**

The evaluation of code summarization is based on three metrics: Completeness, Accuracy, and Readability. Completeness requires the summary to cover six key dimensions: code overview, datasets used, spatial scope, temporal scope, input/output data types, and the functional implementation process. Accuracy demands that, despite potential differences in expression, the semantic content must be correct. Readability focuses on clear logic, smooth phrasing, and concise expression. A prompt was designed to guide GPT-4 to reference the standard answer and assign a score, ranging from 1 to 10, which is then converted to a 0-1 scale for comprehensive evaluation. The prompt design is illustrated in Fig. 6.

Fig.6 Code Summarization Evaluation Prompt Template

As shown in Table 6, GeoCode-GPT-7B demonstrated significant improvements across all three evaluation metrics compared to the baseline model, Code Llama-7B, with gains ranging from 0.142 to 0.252 and an overall enhancement of 0.214 in summarization performance. When compared to other models, GeoCode-GPT-7B showed performance improvements of 0.017 to 0.254 in summarization generation. In the Completeness metric, GeoCode-GPT-7B outperformed all other models by 0.032 to 0.164, except for GPT-4, where it scored 0.026 lower. This difference may be related to context token length, as larger commercial models can provide more contextual information, leading to better completeness scores. All models scored above 0.75, indicating that they could recognize and cover at least 4.5 out of the six dimensions. However, longer context tokens also introduced issues of information redundancy. Although commercial models generated more comprehensive summaries, their efficiency in information expression decreased, resulting in lower scores for Readability. Code generation models, due to their training focus on code, may have diluted abilities in general text generation, with readability scores generally around 0.6. GeoCode-GPT-7B, through expert-optimized instruction tuning with the GeoCode-SFT dataset, led the comparison group in readability, outperforming other models by 0.022 to 0.274.In the Accuracy metric, model performance varied significantly. GPT-4 achieved the highest score at 0.952, while CodeGemma-7B performed the worst at 0.582. GeoCode-GPT-7B, benefiting from instruction fine-tuning that functioned like multiple few-shot learning iterations, improved accuracy by 0.068 to 0.349 compared to other models, though it scored slightly lower than GPT-4 (by 0.022).Overall, the performance of GeoCode-GPT-7B across all summarization metrics validates the effectiveness and advantages of its training strategy.

Table 6 : Code Summarization Evaluation Data Table

| Model | Completeness | | Accuracy | | Readability | | Overall | |
|---|---|---|---|---|---|---|---|---|
| GPT-4 | 0.942 | -0.026 | 0.952 | -0.022 | 0.796 | +0.100 | 0.897 | +0.017 |
| GPT-3.5 | 0.884 | +0.032 | 0.862 | +0.068 | 0.744 | +0.152 | 0.829 | +0.085 |
| ERNIE 4.0 | 0.822 | +0.094 | 0.748 | +0.182 | 0.698 | +0.198 | 0.756 | +0.158 |
| LLaMA 2-7B | 0.780 | +0.136 | 0.698 | +0.232 | 0.852 | +0.044 | 0.777 | +0.137 |
| LLaMA 3-8B | 0.810 | +0.106 | 0.726 | +0.204 | 0.874 | +0.022 | 0.804 | +0.110 |
| CodeGemma-7B | 0.762 | +0.154 | 0.582 | +0.348 | 0.634 | +0.262 | 0.660 | +0.254 |
| StarCoder 2-7B | 0.776 | +0.140 | 0.640 | +0.290 | 0.640 | +0.256 | 0.685 | +0.229 |
| CodeGeeX 2-6B | 0.752 | +0.164 | 0.660 | +0.270 | 0.622 | +0.274 | 0.678 | +0.236 |
| Code Llama-13B | 0.806 | +0.110 | 0.702 | +0.228 | 0.678 | +0.218 | 0.729 | +0.185 |
| Code Llama-7B | 0.774 | +0.142 | 0.678 | +0.252 | 0.648 | +0.248 | 0.700 | +0.214 |
| **GeoCode-GPT-7B** | **0.916** | | **0.930** | | **0.896** | | **0.914** | |

### 5.2.2 Code Generation

The code generation evaluation is based on three core metrics: Accuracy, Readability, and Executability. The scoring methodology is illustrated in the figure. Accuracy refers to the proportion of key entities (such as data sources, time, space, input/output data) in the generated code that match the requirements of the task. This is calculated by using GPT-4 with a specific prompt template to extract and compare the generated code with the target entities, assessing consistency as shown in Fig. 7. Executability is determined by experts who run the generated code on different platforms; the percentage of successfully executed code represents the executability score. Readability evaluates whether the code structure is clear, comments are appropriate, and whether variable naming and function lengths follow standard conventions. Experts rank the generated code from different models through a blind selection process. The readability score is based on the average ranking of each model and is calculated using the formula (11 - n) / 11, where n is the average ranking of the model.

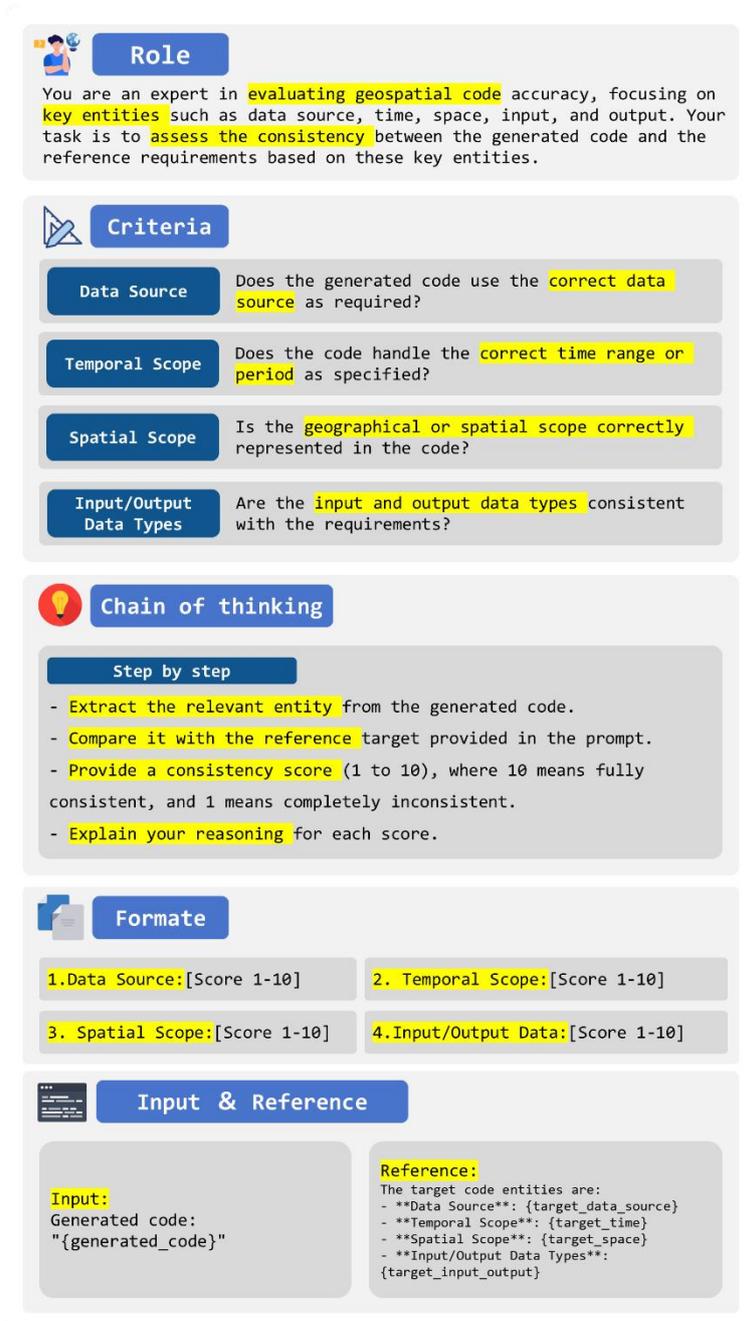

Fig.7 Code Generation Evaluation Prompt Template

As shown in Table 7, GeoCode-GPT-7B demonstrated significant improvements across all metrics compared to the baseline model, particularly in Readability, where the score increased by 0.373. Even in Executability, which showed the smallest improvement, GeoCode-GPT-7B outperformed the baseline by 0.202, with an overall performance increase of 0.251 points. While there remains a gap between GeoCode-GPT-7B and commercial models like GPT-4 in Accuracy, Readability, and Executability (0.036, 0.164, and 0.024, respectively), and a 0.107 gap in Readability compared to GPT-3.5, GeoCode-GPT-7B outperformed other open-source models by 0.08 to 0.373 across all metrics. In terms of Accuracy, general-purpose LLMs and code generation models showed lower accuracy in handling code entities (0.434–0.574), likely due to their lack of geospatial modeling knowledge (e.g., operator syntax and dataset references). Even commercial models failed to exceed an accuracy score of 0.75 in this dimension. Moreover, models without specialized training performed worse in Executability, such as LLaMA 2-7B, which achieved an accuracy of only 0.224. GeoCode-GPT-7B, despite scoring 0.504, still showed an improvement of 0.074 to 0.28 over other models, highlighting the effectiveness of its training strategy. In Readability, there was considerable variation among models. GeoCode-GPT-7B scored 0.698, with an average ranking of 3.322, slightly lower than GPT-4's ranking of 1.529. However, compared to the worst-performing model, Code Llama-7B, GeoCode-GPT-7B improved by 4.103 ranks. Overall, except for a 0.048 gap with GPT-4, GeoCode-GPT-7B outperformed other models by 0.012 to 0.251, demonstrating the potential of fine-tuning for enhancing open-source models in geospatial code generation tasks.

Table 7 : Code Generation Evaluation Data Table

| Model | Accuracy | | Readability | | Executability | | Overall | |
|---|---|---|---|---|---|---|---|---|
| GPT-4 | 0.742 | -0.036 | 0.861 | -0.164 | 0.528 | -0.024 | 0.710 | -0.048 |
| GPT-3.5 | 0.698 | +0.008 | 0.805 | -0.107 | 0.430 | +0.074 | 0.644 | +0.012 |
| ERNIE 4.0 | 0.622 | +0.084 | 0.585 | +0.113 | 0.302 | +0.202 | 0.503 | +0.133 |
| LLaMA 2-7B | 0.434 | +0.272 | 0.623 | +0.074 | 0.224 | +0.280 | 0.427 | +0.209 |
| LLaMA 3-8B | 0.456 | +0.250 | 0.659 | +0.038 | 0.242 | +0.262 | 0.452 | +0.184 |
| CodeGemma-7B | 0.552 | +0.154 | 0.543 | +0.155 | 0.282 | +0.222 | 0.459 | +0.177 |
| StarCoder 2-7B | 0.574 | +0.132 | 0.425 | +0.273 | 0.354 | +0.150 | 0.451 | +0.185 |
| CodeGeeX 2-6B | 0.537 | +0.169 | 0.378 | +0.320 | 0.326 | +0.178 | 0.414 | +0.222 |
| Code Llama-13B | 0.564 | +0.142 | 0.484 | +0.214 | 0.356 | +0.148 | 0.468 | +0.168 |
| Code Llama-7B | 0.528 | +0.178 | 0.325 | +0.373 | 0.302 | +0.202 | 0.385 | +0.251 |
| **GeoCode-GPT-7B** | **0.706** | | **0.698** | | **0.504** | | **0.636** | |

## 6. Conclusion

In this study, we developed the GeoCode-PT and GeoCode-SFT corpora, along with the GeoCode-Eval evaluation set, and introduced GeoCode-GPT-7B, the first large language model dedicated to geospatial code generation. By employing QLoRA and LoRA for pretraining and fine-tuning, we achieved significant improvements in the model's performance. We also established a comprehensive evaluation framework, incorporating option matching, expert validation, and LLM scoring, and conducted a systematic assessment using GeoCode-Eval. The results showed that GeoCode-GPT-7B demonstrated significant improvements over the base model, Code Llama-7B. Although it did not surpass GPT-4 in certain metrics, GeoCode-GPT-7B exhibited clear advantages over models with similar parameter sizes and outperformed larger models such as Code Llama-13B and ERNIE 4.0 in all metrics, as well as GPT-3.5 in most metrics. These results validate the effectiveness of the corpus construction and fine-tuning strategies, as well as GeoCode-GPT-7B's performance superiority in the GeoCode-Eval evaluation.

**6.1 Limitations**

This study provides insights into optimizing the performance boundaries of LLMs in geospatial code generation tasks. To address the gap with GPT-4 in certain metrics, future work will focus on expanding the scale of instruction data, improving data quality, and exploring more efficient methods for structured knowledge representation. Although GeoCode-GPT-7B has shown improvements in code executability, there is still room for enhancement. Future efforts will involve increasing both the scale and precision of high-quality code training corpora, along with leveraging more powerful hardware and additional training epochs to further improve the model's executability in code generation tasks.

**6.2 Outlook**

This study provides a preliminary exploration into the application of LLMs in geospatial modeling tasks. Future research can further unlock their potential by developing models with larger parameter scales and creating customized geospatial code generation models. By integrating multi-platform interfaces, automated code testing could be achieved, reducing reliance on expert knowledge and prompt engineering. Additionally, exploring cross-platform code translation capabilities could lower the syntactic barriers between platforms, enhancing user experience.Further research could focus on models tailored for repository-level geospatial code generation, incorporating indexing, contextual information, and memory modules to handle complex tasks. Lastly, the establishment of multi-agent collaboration frameworks could automate code generation, testing, and optimization, thereby improving both the quality and efficiency of outputs. We anticipate more significant advancements in this field in the near future.

**Reference**

Abdin, M., Jacobs, S.A., Awan, A.A., Aneja, J., Awadallah, A., Awadalla, H., Bach, N., Bahree, A., Bakhtiari, A., Behl, H., 2024. Phi-3 technical report: A highly capable language model locally on your phone. arXiv preprint arXiv:2404.14219. https://doi.org/10.48550/arXiv.2404.14219.

Arievitch, I.M., 2020. Reprint of: The vision of Developmental Teaching and Learning and Bloom's Taxonomy of educational objectives. Learn. Cult. Soc. Interact. 27, 100473. https://doi.org/10.1016/j.lcsi.2020.100473.

Chen, L., Guo, Q., Jia, H., Zeng, Z., Wang, X., Xu, Y., Wu, J., Wang, Y., Gao, Q., Wang, J., 2024. A survey on evaluating large language models in code generation tasks. arXiv preprint arXiv:2408.16498. https://doi.org/10.48550/arXiv.2408.16498.

Chen, M., Tworek, J., Jun, H., Yuan, Q., Pinto, H.P.D.O., Kaplan, J., Edwards, H., Burda, Y., Joseph, N., Brockman, G., 2021. Evaluating large language models trained on code. arXiv preprint arXiv:2107.03374. https://doi.org/10.48550/arXiv.2107.03374.

Coetzee, S., Ivánová, I., Mitasova, H., Brovelli, M.A., 2020. Open geospatial software and data: A review of the current state and a perspective into the future. ISPRS Int. J. Geo-Inf. 9, 90. https://doi.org/10.3390/ijgi9020090.

Cui, J., Li, Z., Yan, Y., Chen, B., Yuan, L., 2023. Chatlaw: Open-source legal large language model with integrated external knowledge bases. arXiv preprint arXiv:2306.16092.

https://doi.org/10.48550/arXiv.2306.16092.

Qin, H., Ma, X., Zheng, X., Li, X., Zhang, Y., Liu, S., Luo, J., Liu, X., Magno, M., 2024. Accurate lora-finetuning quantization of llms via information retention. arXiv preprint arXiv:2402.05445. https://doi.org/10.48550/arXiv.2402.05445.

Granell, C., Díaz, L., Gould, M., 2010. Service-oriented applications for environmental models: Reusable geospatial services. Environ. Modell. Softw. 25, 182-198. https://doi.org/10.1016/j.envsoft.2009.08.005.

Guo, D., Zhu, Q., Yang, D., Xie, Z., Dong, K., Zhang, W., Chen, G., Bi, X., Wu, Y., Li, Y.K., 2024. DeepSeek-Coder: When the Large Language Model Meets Programming--The Rise of Code Intelligence. arXiv preprint arXiv:2401.14196. https://doi.org/10.48550/arXiv.2401.14196.

Hou, S., Zhangxiao, S., Jianyuan, L., Anqi, Z., Zhipeng, G., Rui, L., Wu, H., 2024. Can Large Language Models Generate Geospatial Code? arXiv preprint arXiv:2410.09738. https://doi.org/10.48550/arXiv.2410.09738.

Hu, E.J., Shen, Y., Wallis, P., Allen-Zhu, Z., Li, Y., Wang, S., Wang, L., Chen, W., 2021. Lora: Low-rank adaptation of large language models. arXiv preprint arXiv:2106.09685. https://doi.org/10.48550/arXiv.2106.09685.

Huang, A.H., Wang, H., Yang, Y., 2023. FinBERT: A large language model for extracting information from financial text. Contemp. Account. Res. 40, 806-841. https://doi.org/10.1111/1911-3846.12832.

Jiang, J., Wang, F., Shen, J., Kim, S., Kim, S., 2024. A Survey on Large Language Models for Code Generation. arXiv preprint arXiv:2406.00515. https://doi.org/10.48550/arXiv.2406.00515.

Kasneci, E., Seßler, K., Küchemann, S., Bannert, M., Dementieva, D., Fischer, F., Gasser, U., Groh, G., Günnemann, S., Hüllermeier, E., 2023. ChatGPT for good? On opportunities and challenges of large language models for education. Learn. Individ. Differ. 103, 102274. https://doi.org/10.1016/j.lindif.2023.102274.

Kifetew, F.M., Tiella, R., Tonella, P., 2017. Generating valid grammar-based test inputs by means of genetic programming and annotated grammars. Empir. Softw. Eng. 22, 928-961. https://doi.org/10.1007/s10664-015-9422-4.

Li, L., Jiang, X., Liu, Q., 2019. Pretrained language models for document-level neural machine translation. arXiv preprint arXiv:1911.03110. https://doi.org/10.48550/arXiv.1911.03110.

Lirov, Y., 1991. Computer-aided software engineering of expert systems. Expert Syst. Appl. 2, 333-343. https://doi.org/10.1016/0957-4174(91)90039-H.

Luo, Z., Xu, C., Zhao, P., Sun, Q., Geng, X., Hu, W., Tao, C., Ma, J., Lin, Q., Jiang, D., 2023. Wizardcoder: Empowering code large language models with evol-instruct. arXiv preprint arXiv:2306.08568. https://doi.org/10.48550/arXiv.2306.08568.

Minaee, S., Mikolov, T., Nikzad, N., Chenaghlu, M., Socher, R., Amatriain, X., Gao, J., 2024. Large language models: A survey. arXiv preprint arXiv:2402.06196. https://doi.org/10.48550/arXiv.2402.06196.

Nymeyer, A., Katoen, J.P., 1997. Code generation based on formal BURS theory and heuristic search. Acta Inform. 34, 597-635. https://doi.org/10.1007/s002360050099.

Roziere, B., Gehring, J., Gloeckle, F., Sootla, S., Gat, I., Tan, X.E., Adi, Y., Liu, J., Sauvestre, R., Remez, T., 2023. Code llama: Open foundation models for code. arXiv preprint arXiv:2308.12950. https://doi.org/10.48550/arXiv.2308.12950.

Scalabrino, S., Bavota, G., Vendome, C., Linares-Vasquez, M., Poshyvanyk, D., Oliveto, R., 2019. Automatically assessing code understandability. IEEE Trans. Softw. Eng. 47, 595-613.

https://doi.org/10.1109/TSE.2019.2901468.

Thirunavukarasu, A.J., Ting, D.S.J., Elangovan, K., Gutierrez, L., Tan, T.F., Ting, D.S.W., 2023. Large language models in medicine. Nat. Med. 29, 1930-1940. https://doi.org/10.1038/s41591-023-02448-8.

Tonmoy, S.M., Zaman, S.M., Jain, V., Rani, A., Rawte, V., Chadha, A., Das, A., 2024. A comprehensive survey of hallucination mitigation techniques in large language models. arXiv preprint arXiv:2401.01313. https://doi.org/10.48550/arXiv.2401.01313.

Wang, J., Zhang, B., Du, Q., Zhang, J., Chu, D., 2024. A Survey on Data Selection for LLM Instruction Tuning. arXiv preprint arXiv:2402.05123. https://doi.org/10.48550/arXiv.2402.05123.

Wang, Y., Kordi, Y., Mishra, S., Liu, A., Smith, N.A., Khashabi, D., Hajishirzi, H., 2022. Self-instruct: Aligning language models with self-generated instructions. arXiv preprint arXiv:2212.10560. https://doi.org/10.48550/arXiv.2212.10560.

Wu, S., Irsoy, O., Lu, S., Dabravolski, V., Dredze, M., Gehrmann, S., Kambadur, P., Rosenberg, D., Mann, G., 2023. Bloomberggpt: A large language model for finance. arXiv preprint arXiv:2303.17564. https://doi.org/10.48550/arXiv.2303.17564.

Zhang, X., Tian, C., Yang, X., Chen, L., Li, Z., Petzold, L.R., 2023. Alpacare: Instruction-tuned large language models for medical application. arXiv preprint arXiv:2310.14558. https://doi.org/10.48550/arXiv.2310.14558.

Zhang, Y., Wang, Z., He, Z., Li, J., Mai, G., Lin, J., Wei, C., Yu, W., 2024a. BB-GeoGPT: A framework for learning a large language model for geographic information science. Inf. Process. Manage. 61, 103808. https://doi.org/10.1016/j.ipm.2024.103808.

Zhang, Y., Wei, C., He, Z., Yu, W., 2024b. GeoGPT: An assistant for understanding and processing geospatial tasks. Int. J. Appl. Earth Obs. Geoinf. 131, 103976. https://doi.org/10.1016/j.jag.2024.103976.

Zhao, W.X., Zhou, K., Li, J., Tang, T., Wang, X., Hou, Y., Min, Y., Zhang, B., Zhang, J., Dong, Z., 2023. A survey of large language models. arXiv preprint arXiv:2303.18223. https://doi.org/10.48550/arXiv.2303.18223.

Wei, H., He, S., Xia, T., Wong, A., Lin, J., Han, M., 2024. Systematic Evaluation of LLM-as-a-Judge in LLM Alignment Tasks: Explainable Metrics and Diverse Prompt Templates. arXiv preprint arXiv:2408.13006. https://doi.org/10.48550/arXiv.2408.13006.

Zhou, C., Liu, P., Xu, P., Iyer, S., Sun, J., Mao, Y., Ma, X., Efrat, A., Yu, P., Yu, L., 2023. LIMA: Less Is More for Alignment. arXiv preprint arXiv:2305.11206. https://doi.org/10.48550/arXiv.2305.11206.

Zhuo, T.Y., 2023. ICE-Score: Instructing Large Language Models to Evaluate Code. arXiv preprint arXiv:2304.14317. https://doi.org/10.48550/arXiv.2304.14317.